\documentstyle[12pt]{article}

\begin{document}

\begin{center}
{\Large\bf Massless fields in plane wave geometry}
\end{center}

\vspace{2cm}
\centerline{R.R. Metsaev}

\vspace{1cm}
\begin{center}
{\it Department of Theoretical Physics, P.N. Lebedev Physical
Institute, 117924, Leninsky prospect 53, Moscow, Russia.}
\end{center}

\vspace{3cm}
\begin{center}
{\bf Abstract}
\end{center}

\bigskip

Conformal isometry algebras of plane wave geometry are studied.
Then, based on the requirement of conformal invariance,
a definition of masslessness is introduced and gauge
invariant equations of motion, subsidiary conditions, and
corresponding gauge transformations for all plane wave
geometry massless spin fields are constructed.  Light cone
representation for elements of conformal algebra acting as
differential operators on wavefunctions of massless higher spin
fields is also evaluated.  Interrelation of plane wave geometry
massless higher spin fields with ladder representation of
$u(2,2)$ algebra is investigated.


\newpage

\section{Introduction}

The plane wave \cite{1} has been actively studied from
various point of view in the context of both field theory (see
\cite{BPR} \cite{Penrose}  \cite{Deser} \cite{Gibbons}) and
string theory (see \cite{NW} \cite{HORST} \cite{HORTS} \cite{TS1}
\cite{KEHM} \cite{AOBERS} \cite{FHHP}).  One of the remarkable
properties of the plane wave is that it provides a non-perturbative
solution to the classical equations of motion of superstring
effective field theory. It is this property that has inspired a
renewed interest in the plane wave in recent years. At
present time a significant role of the plane wave is beyond doubt.
Due
to that, it is highly desirable to study the questions related to
fields propagating in plane wave geometry (see \cite{Gibbons}).

In this paper we study the following questions
related to massless fields: (1) definition of
masslessness in four-dimensional plane wave geometry; (2) gauge
invariant equations of motion and corresponding gauge
transformations as well as subsidiary conditions; (3) relation of
plane wave geometry massless fields to ladder representations.

The paper is organized as follows.
In Sec. II we shall review isometry algebras of various plane
wave backgrounds and construct complete expressions
for generators, i.e., corresponding orbital and spin
parts.
In Sec. III we shall consider conformal isometry
algebras.  We present complete manifest expressions for
elements of conformal algebras, i.e., corresponding
orbital, spin, and conformal boost parts.  In Sec. IV, based on
conformal invariance, we shall construct free equations of
motion for all, lower as well as higher, massless spin fields.
These equations we supplement by corresponding gauge
transformations and subsidiary conditions.  We study the wave
equation of motion for massless field and demonstrate an
existence a nonzero energy lowest value for a massless field.  We
construct also a light cone representation for elements of
a conformal algebra acting as differential operators on a
wavefunction of massless field.  In Sec. V we shall demonstrate
explicitly the manner in which the infinity chain of plane wave
geometry massless spin fields
$\lambda =0, \pm1/2, \pm1, \pm3/2,\pm2,\ldots$
(every state appears once) can be embedded into
ladder representation of the $u(2,2)$ algebra.  By using ladder
representations we demonstrate also how Minkowski and
plane wave geometry massless fields are related to each other.
In Sec. VI we shall describe possible applications of our
results.  Appendices A and B detail certain mathematical
manipulations.

\section{Isometry Algebras}

\subsection{General case}
The metric for a general gravitational-electromagnetic
plane wave can be read as follows:

\begin{equation}\label{metric}
ds^2=2dudv+2(f(u)\zeta^2+\bar{f}(u)\bar{\zeta}^2
+F(u)\zeta\bar{\zeta})du^2-2d\zeta d\bar{\zeta}\,.
\end{equation}
Note that (\ref{metric}) is sometimes called an exact plane
wave.  The metric (\ref{metric}) is written in coordinates
$(u,v,\zeta,\bar{\zeta})$.  In what follows it will be
convenient to use the coordinates $x^\mu=(u,v,\zeta_i)$, where
$\zeta_i$ are given by

\begin{equation}\label{complex}
\zeta=(\zeta_1+{\rm i}\zeta_2)/\sqrt{2}\,,\qquad
\bar{\zeta}=(\zeta_1-{\rm i}\zeta_2)/\sqrt{2}\,.
\end{equation}
All the variables $u$, $v$, and $\zeta_i$ range from $-\infty$ to
$+\infty$.
Throughout this paper, unless otherwise specified, we use the
following notation and conventions: (1) the indices $i,j$, and
$k$ run over 1,2; (2) $\zeta^2$, $\partial^2$, and
$(\zeta\partial)$ stand for $\zeta_i\zeta_i$,
$\partial_i\partial_i$, and $\zeta_i\partial_i$ respectively,
where $\partial_i=\partial/\partial \zeta^i$, $\zeta_i=\zeta^i$.

In Ref.\cite{AOBERS} (see also Ref.\cite{BPR})
the various plane waves
have been classified according to the number of Killing
symmetries they possess:

$$
\begin{array}{cccc}
 &\quad\quad f(u)\quad\quad &\quad F(u) \quad
& \hbox{dim (Killing sym)}\\[5pt]
I  & 0  & F={\kern-2ex/}\,\,0  &  7\\[5pt]
II & 0  & F(u)                 &  6 \\[5pt]
III& f={\kern-2ex/}\,\,0   & F   &  6 \\[5pt]
IV & f(u)                 & F(u) & 5  \\[5pt]
\end{array}
$$

\noindent
where $f$ and $F$ without argument stand for $u$-independent
constants. Note that the I and II cases (vanishing Weyl
tensor) are usually called purely electromagnetic plane wave,
while one of the particulars of III and IV when $F(u)=0$
(vanishing Ricci tensor) is called purely gravitational plane
wave.  The Killing vectors for $I-IV$ can be read as
(see also Refs.\cite{AOBERS} and \cite{FHHP})

$$
\begin{array}{ccccc}
\qquad I\qquad   &\quad   P_v\,,\quad
  &\quad P_C\,,\quad &\quad P_u\,,\quad & J_{ij}\,,\\[5pt]
\qquad II \qquad  &   P_v\,,      &P_C\,,   &  &J_{ij}\,, \\[5pt]
\qquad III\qquad &   P_v\,,      &P_C\,,  & P_u\,, \\[5pt]
\qquad IV\qquad   &   P_v\,,      &P_C\,,
\end{array}
$$
where
\begin{equation}\label{gen1}
P_v=\partial_v\,,\quad P_u=\partial_u\,,
\end{equation}

\begin{equation}\label{gen2}
l(P_C)=C_i^\prime\zeta_i\partial_v
+C_i \partial_i\,,
\end{equation}
\begin{equation}\label{gen3}
l(J_{ij})=-\zeta_i\partial_j+\zeta_j\partial_i\,,
\end{equation}
and prime indicates partial derivative with respect to $u$.
Here $C_i$ satisfy the following second-order differential
equation:
\begin{equation}\label{nssoe}
C^{\prime\prime}+2\bar{f}(u)\bar{C}+F(u)C=0\,,
\end{equation}
where $C=(C_1+{\rm i}C_2)/\sqrt{2}$, $\bar{C}=(C_1-{\rm
i}C_2)/\sqrt{2}$. In (\ref{gen2}),(\ref{gen3}) and below a
notation $l(G)$ is used to indicate the fact that corresponding
expression on the rhs provides only the orbital part of algebra
element $G$.

\subsection{I and II cases}

The I and II cases have a rotation
Killing vector (\ref{gen3}). Due to that we are able to
introduce spin-$s$ tensor fields $\Phi^{\mu_1,\ldots,\mu_s}$,
where the spin $s$ can be associated with eigenvalue of $J_{ij}$.
Since the expressions (\ref{gen1})-(\ref{gen3}) provide
only orbital parts they should be accompanied by
corresponding spin parts.  To derive spin parts we proceed as
follows. First of all, we would like to pass to
tangent space tensor fields. To do that we should introduce local
a frame. One convenient choice is specified by the frame
one-forms ${\bf e}^A={\bf e}^A_\mu dx^\mu$ with

$$
ds^2=2{\bf e}^U{\bf e}^V-{\bf e}^i{\bf e}^i\,,
$$
\begin{eqnarray}\label{1forms}
&& {\bf e}^U=du\,, \nonumber \\
&& {\bf e}^V=dv+\frac{1}{2}F(u)\zeta^2du \,,\\
&& {\bf e}^i=d\zeta^i\,. \nonumber
\end{eqnarray}
The connection one-forms, defined by
$$
d{\bf e}^A+\omega^A{}_B\wedge {\bf e}^B=0\,,
\qquad \omega^{AB}=-\omega^{BA}
$$
are then given by

\begin{equation}\label{connec}
\omega^{Vi}=-F(u)\zeta^i e^U\,.
\end{equation}
Dual forms $e_{_A}$ to (\ref{1forms}),
$e_{_A}({\bf e}^B)=\delta_A^B$, read as

\begin{eqnarray}\label{dforms}
&&e_{_U}=\partial_u-\frac{1}{2}F(u)\zeta^2\partial_v\,,
\nonumber\\ && e_{_V}=\partial_v\,,\\ && e_i=\partial_i \,.
\nonumber \end{eqnarray} Now we introduce the tangent space
tensor fields in the usual manner:

\begin{equation}\label{trans}
\Phi^{A_1\ldots A_s}
={\bf e}_{\mu_1}^{A_1}\ldots {\bf e}_{\mu_s}^{A_s}
\Phi^{\mu_1 \ldots \mu_s}\,.
\end{equation}
Then to avoid cumbersome tensor expressions
we introduce creation and annihilation operators $a_{_A}$ and
$\bar{a}_{_A}$ which satisfy

\begin{equation}\label{osccr}
[\bar{a}_{_A},\,a_{_B}]=-\eta_{_{AB}}\,,
\qquad
\eta_{_{UV}}=\eta_{_{VU}}=1\,,
\quad
\eta_{ij}=-\delta_{ij}\,,
\end{equation}
and construct a Fock space vector

\begin{equation}\label{genfun}
|\Phi\rangle
=a_{_{A_1}}\ldots a_{_{A_s}}
\Phi^{{\scriptstyle A_1\ldots A_s}}|0\rangle\,,
\qquad \bar{a}_{_A}|0\rangle=0\,.
\end{equation}
Now the nonvanishing spin parts of isometry algebra under
consideration are given by

\begin{eqnarray}
&&
s(J_{ij})=M_{ij}\,,\nonumber\\
\label{spinp1}
&&         \\ [-10pt]
&&
s(P_C)=C_i^\prime M_{{\scriptscriptstyle V}i}\,,
\nonumber
\end{eqnarray}
where the elements $M_{AB}$ constitute $so(3,1)$ algebra

\begin{equation}\label{Lorenal}
[M_{_{AB}},\,M_{_{CD}}]=\eta_{_{AC}}M_{_{DB}}+\ldots
\end{equation}
(for details see Appendix A).

\setcounter{equation}{0}

\section{Conformal Isometry Algebras}

The purpose of this section is to get a manifest representation
for elements of a conformal isometry algebra of plane wave
geometry.  This representation will be used throughout the
paper. Before moving to the details, let us comment on the
I and II cases.  Since for these cases the Weyl tensor vanishes,
the conformal algebra in question is isomorphic to the $so(4,2)$
algebra. For these cases we also establish the isomorphism
explicitly, i.e., we take each element of the conformal algebra
calculated to an element of $so(4,2)$ algebra taken in the
Lorentz basis.

As usual we solve the equations for conformal Killing vectors

\begin{equation}\label{conkileq}
\nabla_\mu \xi_\nu+\nabla_\nu \xi_\mu
=\frac{1}{2}g_{\mu\nu}\nabla_\rho \xi^\rho\,.
\end{equation}
The result of the solution can be summarized as follows:

$$
\begin{array}{clllllll}
I   &  P_v\,,\quad &P_C\,,\quad &J_{ij}\,,\quad
    &  D\,,     &T_A\,,               &K_C\,, & K_u\,,\\[5pt]
II  &  P_v\,,\quad &P_C\,,      &J_{ij}\,,
    &  D\,,\quad     &T_A\,,        &K_C\,,& K_u\,, \\[5pt]
III &  P_v\,,\quad &P_C\,,\quad
    &   &D\,,  &P_u\,,&&   \\[5pt]
IV  &   P_v\,,\quad &P_C\,,&  &D\,,&&&
\end{array}
$$
where

\begin{eqnarray}
&&
l(T_A)=\frac{1}{4}\zeta^2 A^{\prime\prime} \partial_v
+A\partial_u+\frac{1}{2}A^\prime(\zeta\partial)\,,
\nonumber \\ [5pt]
&&
l(K_C)=\zeta_i(v C_i^{\prime}
-\frac{1}{2}\zeta^2 F(u)C_i)\partial_v
+C_i\zeta_i \partial_u
+C_i^\prime \zeta_i(\zeta\partial)
+(v C_i- \frac{1}{2}\zeta^2 C_i^\prime)\partial_i\,,
\nonumber\\
\label{congen1}&&  \\ [-10pt]
&&
l(K_u)=(v^2-\frac{1}{4}F(u)(\zeta^2)^2)\partial_v
+\frac{1}{2}\zeta^2\partial_u
+v(\zeta\partial)\,,
\nonumber \\ [5pt]
&&
l(D)=2v\partial_v +(\zeta\partial)\,.
\nonumber
\end{eqnarray}
The functions $C_i$ satisfy the equation (\ref{nssoe}) which for
the $I$ and $II$ cases can be rewritten as follows:

\begin{equation}\label{soe}
C_i^{\prime\prime}+F(u)C_i=0\,.
\end{equation}
The function $A$ is relevant only for the $I$
and $II$ cases and it satisfies the following third-order
differential equation:

\begin{equation}\label{toe}
A^{\prime\prime\prime}+4F(u)A^\prime+2F^\prime(u)A=0\,.
\end{equation}
Note that for the I case the isometry element $P_u$ is included
into $T_A$.

Now let $y_\alpha$, $\alpha=1,2$, and $Y_a$, $a=1,2,3$,
be independent solutions of the equation
\begin{equation}\label{soesc}
y^{\prime\prime}+F(u)y=0
\end{equation}
and (\ref{toe}), respectively.  Consider the following brackets:

\begin{eqnarray}
&& \{Y_a,\,y_\beta\}\equiv Y_a y_\beta^\prime
   -\frac{1}{2} y_\beta Y_a^\prime\,,
\nonumber  \\ [3pt]
\label{com1}
&& \{ y_\alpha,\,y_\beta\}\equiv y_\alpha y_\beta\,,
\\ [3pt]
&& \{ Y_a,\, Y_b\}\equiv Y_a Y_b^\prime
-Y_b Y_a^\prime \,.      \nonumber
\end{eqnarray}
It is easy to prove that $\{Y_a,\,y_\beta\}$ and
$\{ y_\alpha,\,y_\beta\}$,
$\{ Y_a,\, Y_b\}$ satisfy the
equations (\ref{soesc}) and (\ref{toe}), respectively. In other
words, the following decompositions hold true:

\begin{eqnarray}
&&
{} \{Y_a,\,y_\beta\}
=f_{a\beta}^\alpha y_\alpha\,,
\nonumber\\
\label{supa1} &&
{} \{ y_\alpha,\,y_\beta\}=f_{\alpha\beta}^a Y_a\,,
\\
&&
\{ Y_a,\, Y_b\}
=f_{ab}^c Y_c\,.
\nonumber
\end{eqnarray}
An interesting fact is that the $y_\alpha$ and
$Y_a$ equipped
with commutation relations (\ref{supa1}) constitute
the superalgebra $osp(2,1)$ (for proof see Appendix B). Now we
are able to write down the commutation relations of our algebra:

\begin{equation}\label{alcom-1}
{} [D,\,P_C]=-P_C\,,
\quad
{} [D,\,P_v]=-2P_v\,,
\quad
{}[D,\,K_C]=K_C\,,
\quad
{} [D,\,K_u]=2K_u\,,
\end{equation}

\begin{equation}\label{alcom1}
{}
[P_{C},\,P_{B}]=W(C_i,B_i)P_v\,,\qquad
[K_{C},\,K_{B}]=W(C_i,B_i)K_u\,,
\end{equation}

\begin{equation}\label{alcom02}
{} [T_{Y_1},\,T_{Y_2}]=T_{\{Y_1,\,Y_2\}}\,,
\end{equation}

\begin{equation}\label{alcom2}
[T_Y,\, P_C]=
P_{\{Y,\, C\}}\,,
\end{equation}

\begin{equation}\label{alcom3}
[T_Y,\, K_C]=
K_{\{Y,\, C\}}\,,
\end{equation}

\begin{equation}\label{alcom4}
[K_u,\, P_C]=-K_C\,,
\end{equation}

\begin{equation}\label{alcom5}
[P_v,\,K_C]=P_C\,,
\end{equation}

\begin{equation}\label{alcom6}
[P_C,\, K_B]=
T_{\{C_i,\,B_i\}}
+W(C_i,B_j)(\frac{1}{2}\delta_{ij}D
+J_{ij})\,,
\end{equation}
where we introduce

$$
W(C_i,B_j)\equiv C_i B_j^\prime- B_j C_i^\prime\,.
$$
Note that due to eq.(\ref{soe}), the $W$ does not
depend on $u$.

Now we are going to prove that the algebra of commutators of
(\ref{alcom1})-(\ref{alcom6}) is isomorphic to that of the usual
conformal $so(4,2)$ algebra. To do that we prefer to use the
`tensor' form of the algebra under consideration. First of all,
let us write the general solution to (\ref{soe}) as follows:

$$
C_i=C_{0i}^+y^++C_{0i}^-y^-\,,\qquad
y^\pm = (y_1\pm {\rm i} y_2)/\sqrt{W}\,,
$$
where $C_{0i}^\pm$ are some constants while  $y_1$ and $y_2$
are real-valued solutions to  (\ref{soesc}). For definiteness
we consider the $W>0$ case.  Note that three independent
solutions to (\ref{toe}) can be written in terms of $y^\pm$ as
follows:

$$
Y^0=y^+y^-\,,\qquad Y^{\pm\pm} = (y^\pm)^2\,.
$$
Then we introduce desired generators

$$
P_i^\pm=\frac{\partial}{\partial C_{0i}^\pm}P_C\,,
\qquad
K_i^\pm=\frac{\partial}{\partial C_{0i}^\pm}K_C\,,
$$
$$
P_u\equiv T_{Y^0}\,,\qquad
J^{\pm\pm}\equiv T_{Y^{\pm\pm}}\,.
$$
Commutation relations between these elements can be readily
derived from those of (\ref{alcom1}-\ref{alcom6}).
The elements $\{D,P_v,K_u\}$, $\{P_u,J^{\pm\pm}\}$, and
$J_{ij}$ form three commutative subalgebras
$K\equiv so(2,1)\otimes so(2,1)\otimes so(2)$, respectively:

\begin{equation}\label{kk1}
{} [D,\,P_v]=-2P_v\,,\quad
{} [D,\,K_u]=2K_u\,,\quad
{} [P_v,\,K_u]=D\,,
\end{equation}

\begin{equation}\label{kk2}
{} [J^{\pm\pm},\,P_u]=\pm 2 {\rm i}J^{\pm\pm}\,,
\quad
{} [J^{++},\,J^{--}]=4{\rm i}P_u\,.
\end{equation}
All remainder elements, which are $V\equiv P_i^\pm$, $K_i^\pm$,
transform in their representations $[V,\,K]\in V$:

\begin{equation}\label{vk1}
{} [D,\,P_i^\pm]=-P_i^\pm\,,
\quad
{} [D,\,K_i^\pm]=K_i^\pm\,,
\end{equation}

\begin{equation}\label{vk2}
{} [P_i^\pm,\, K_u]=K_i^\pm\,,
\quad
{} [K_i^\pm,\, P_v]=-P_i^\pm\,,
\end{equation}

\begin{equation}\label{vk3}
{} [P_i^\pm,\,P_u]=\pm {\rm i}P_i^\pm\,,
\quad
{} [K_i^\pm,\,P_u]=\pm {\rm i}K_i^\pm\,,
\end{equation}

\begin{equation}\label{vk4}
{} [J^{\pm\pm},\, P_i^\mp]=\pm 2 {\rm i} P_i^\pm\,,
\quad
{} [J^{\pm\pm},\, K_i^\mp]=\pm 2  {\rm i} K_i^\pm\,,
\end{equation}

\begin{equation}\label{vk5}
{} [P_i^\pm,\,J_{jk}]=\delta_{ij}P_k^\pm-\delta_{ik}P_j^\pm\,,
\qquad
{} [K_i^\pm,\,J_{jk}]=\delta_{ij}K_k^\pm-\delta_{ik}K_j^\pm\,,
\end{equation}
and satisfy the following commutation relations between each
other $[V,\,V]\in K$:

\begin{equation}\label{vv1}
{} [P_i^+,\,P_j^-]=2{\rm i}\delta_{ij}P_v\,,\quad
{} [K_i^+,\,K_j^-]=2{\rm i}\delta_{ij}K_u\,,
\end{equation}

\begin{equation}\label{vv2}
{} [P_i^\pm,\, K_j^\mp]=\delta_{ij}P_u
\pm{\rm i}(\delta_{ij}D+2J_{ij})\,,
\end{equation}

\begin{equation}\label{vv3}
{} [P_i^\pm,\,K_j^\pm]=\delta_{ij}J^{\pm\pm}\,.
\end{equation}
In other words, the algebra
under consideration has the Cartan-like decomposition

$$
G=V
{\,\lower-0.2ex\hbox{${\scriptstyle+}$}}
{\kern-1.3ex\hbox{$\supset$}\,}
K\,.
$$
Before formulating our statement, let us decompose
$P_i^\pm$ and $K_i^\pm$ into real pieces

\begin{equation}\label{realdec}
P_i^\pm=P_i\mp {\rm i} J_{vi}\,,
\end{equation}

\begin{equation}\label{realdec2}
K_i^\pm=\pm {\rm i}K_i+J_{ui}\,.
\end{equation}
Now we are in a position to show explicitly an
isomorphism between the algebra above and the elements of
conformal $so(4,2)$ algebra taken in the Lorentz basis.  Our
statement is that the generators defined by

\begin{eqnarray}
{\bf P}_u&=&\frac{1}{2}P_u+\frac{1}{4}(J^{++}+J^{--})\,,
\nonumber \\[5pt]
{\bf K}_v&=&\frac{1}{2}P_u-\frac{1}{4}(J^{++}+J^{--})\,,
\nonumber \\
\label{link} && \\ [-5pt]
{\bf D}&=&\frac{1}{2}D+\frac{{\rm i}}{4}(J^{++}-J^{--})\,,
\nonumber \\[5pt]
{\bf J}_{uv}&=&\frac{1}{2}D-\frac{{\rm i}}{4}(J^{++}-J^{--})\,,
\nonumber
\end{eqnarray}

$$
{\bf P}_v=P_v\,,\quad {\bf P}_i=P_i\,,\quad
{\bf K}_u=K_u\,,\quad {\bf K}_i=K_i\,,
$$

$$
{\bf J}_{vi}=J_{vi}\,,\quad
{\bf J}_{ui}=J_{ui}\,,\quad
{\bf J}_{ij}=J_{ij}\,,
$$
satisfy the familiar commutation relations of the $so(4,2)$
algebra

$$
\begin{array}{l}
{} [{\bf J}_{\mu\nu},\,{\bf J}_{\rho\sigma}]=
\eta_{\mu\rho}{\bf J}_{\sigma\nu}+\ldots\,,
\\ [5pt]
{} [{\bf P}_\mu,\,{\bf J}_{\rho\sigma}]=
\eta_{\mu\rho}{\bf P}_\sigma+\ldots\,,
\\ [5pt]
{} [{\bf K}_\mu,\,{\bf J}_{\rho\sigma}]=
\eta_{\mu\rho}{\bf K}_\sigma+\ldots\,,
\end{array}\qquad
\begin{array}{l}
{} [{\bf D},\,{\bf P}_\mu]=-{\bf P}_\mu\,,
\\ [5pt]
{} [{\bf D},\,{\bf K}_\mu]={\bf K}_\mu\,,
\\   [5pt]
{} [{\bf D},\,{\bf J}_{\mu\nu}]=0\,,
\end{array}
$$

$$
[{\bf P}_\mu,\,{\bf K}_\nu]=\eta_{\mu\nu}{\bf D}-{\bf J}_{\mu\nu}\,,
\qquad
[{\bf P}_\mu,\,{\bf P}_\nu]=0\,,\qquad
[{\bf K}_\mu,\,{\bf K}_\nu]=0\,,
$$
where nonvanishing elements of $\eta_{\mu\nu}$ are given by
$\eta_{uv}=\eta_{vu}=1$ and $\eta_{ij}=-\delta_{ij}$. Making
use of (\ref{kk1}-\ref{realdec2}), this statement can be easily
proved. For future reference let us describe how we link
generators above with those taken in the six dimensional frame.
We introduce $J_{AB}$, $A,B=(\mu,5,6)$, as

\begin{eqnarray}
&&
{\bf P}_\mu=(J_{5\mu}+J_{6\mu})/\sqrt{2}\,,\qquad
{\bf K}_\mu=(J_{5\mu}-J_{6\mu})/\sqrt{2}\,,
\nonumber \\ [-3pt]
\label{Lorsix}&& \\
&& {\bf J}_{\mu\nu}=J_{\mu\nu}\,,\qquad
{\bf D}=J_{56}
\nonumber
\end{eqnarray}
which satisfy the well-known commutation relations
$$
{} [J_{AB},\, J_{CD}]=
\eta_{AC}^{\vphantom{5pt}}J_{DB}+\ldots,
$$
where
$\eta_{uv}=\eta_{vu}=\eta_{55}=-\eta_{11}
=-\eta_{22}=-\eta_{66}=1$.

Now we would like to write down the expressions for the spin parts
of the conformal algebra elements.
The procedure of evaluation is the same that we used for the
spin parts of isometry algebra (see Appendix A).
The result is given by

\begin{eqnarray}
&&
s(T_A)=\frac{1}{2}A^{\prime\prime}\zeta_iM_{Vi}
+\frac{1}{2}A^\prime(\Delta+2M_{VU})\,,
\nonumber \\ [5pt]
&&
s(K_C)=(vC_i^\prime-F(u)\zeta_i (\zeta C))M_{Vi}
+C_i M_{Ui}
+(\zeta C^\prime)(\Delta+M_{VU})
+\zeta_i C_j^\prime M_{ij}\,,
\nonumber\\[2pt]
\label{k1} && \\ [-10pt]
&&
s(K_u)=-\frac{1}{2}F(u)\zeta_i\zeta^2 M_{Vi}
+\zeta_iM_{Ui}
+v\Delta\,,
\nonumber\\
&& s(D)=\Delta\,.\nonumber
\end{eqnarray}
Let us comment on the expressions above. As is seen from
(\ref{gen1})-(\ref{gen3}) and (\ref{congen1}) the
little algebra which leaves the point $x=0$ invariant is
given by eleven elements that are six homogeneous Lorentz
transformations $M_{AB}$, dilatation $\Delta$, and special
conformal transformations
$\kappa_{_A}=(\kappa_{_U}$,$\kappa_{_V}$,$\kappa_i$).
The expressions (\ref{k1}) include only $M_{_{AB}}$ and
$\Delta$ parts.
As usual, the $\kappa_{_A}$ part of the conformal algebra can
be obtained by considering the homogeneous $so(4,2)$
transformations.  The result is given by

\begin{eqnarray*}
&&
k(T_A)=\frac{1}{2}A^{\prime\prime}\kappa_{_V}\,,
\\ [5pt]
&&
k(K_C)=-F(u)(C\zeta)\kappa_{_V}-(C^\prime \kappa)\,,
\\ [5pt]
&&
k(K_u)=\kappa_{_U}-\frac{1}{2}F(u)\zeta^2\kappa_{_V}\,,
\end{eqnarray*}
where $\kappa_{_A}$ transforms in vector representation
of the $so(3,1)$ algebra:

$$
[\kappa_{_A},\,M_{_{BC}}]=\eta_{_{AB}}\kappa_{_C}-
\eta_{_{AC}}\kappa_{_B}\,.
$$
Note that $[\Delta,\,\kappa_A]=0$, while the commutator
$[\Delta,\,M_{AB}]$ can be read from (A.1), (A.2), and
(\ref{osccr}). Finally, the complete expressions for conformal
generators read as \begin{equation}\label{oskdec}
G=l(G)+s(G)+k(G)\,.
\end{equation}

\setcounter{equation}{0}

\section{Massless Fields}

\subsection{Gauge invariant equations of motion}

With conformal algebra at our hands we are ready to provide a
constructive definition of massless fields living in
plane wave geometry. Notice from now on we
restrict ourselves to the I case because for this case only
there exists a conserved energy Killing vector. By analogy with
Minkowski and de Sitter spaces we define plane wave geometry
massless fields as those whose wave equations are conformal
invariant.  Now let us construct relevant wave equations.

First of all let us collect results concerning the manifest
expressions for isometry algebra elements. From now on
to simplify our expressions we shall put $F=1$. Then making use
(\ref{gen2}) and (\ref{soe}) one gets

\begin{equation}\label{difreal1}
P_u=\partial_u\,,\qquad
P_v=\partial_v\,,
\end{equation}

\begin{equation}\label{difreal2}
P_i^\pm=e^{\mp {\rm i}u}_{\vphantom{5pt}}
(\partial_i\mp {\rm i}\zeta_i\partial_v
\mp {\rm i} M_{Vi})\,,
\end{equation}

\begin{equation}\label{difreal3}
J_{ij}=-\zeta_i\partial_j+\zeta_j\partial_i+M_{ij}\,.
\end{equation}
From the commutation relations which satisfy the generators
above (see [\ref{vk5} and \ref{vv1}]) it is readily seen that the
algebra under consideration has the following Levy
decomposition:

$$
G=N
{\,\lower-0.2ex\hbox{${\scriptstyle+}$}}
{\kern-1.3ex\hbox{$\supset$}\,}
S\,,
$$
where $S$  is a `semisimple' algebra $so(2)$ spanned by $J_{ij}$
while $N$ is a maximal solvable ideal (radical) spanned by
$P_u$, $P_v$, $P_i^\pm$. Note that the $N$ is
nothing but a six dimensional Heisenberg algebra $H_6$. This fact
can be readily seen from (\ref{vk3}) and (\ref{vv1}). As
is known, the Killing metric for the $H_6$ is degenerate.
By analogy with the Poincare algebra one might replace it by a
certain invariant symmetric nondegenerate bilinear form. In
turns out that there exist two forms: the first form leads to the
following second-order operator,

\begin{equation}\label{casimir2}
Q=2P_uP_v-\frac{1}{2}P_i^+P_i^--\frac{1}{2}P_i^-P_i^+\,,
\end{equation}
and it is an analog of $p^2$, while the second form leads to

\begin{equation}\label{heloper}
Q_\lambda
=\epsilon_{ij}J_{ij}P_v+{\rm i}\epsilon_{ij}P_i^+P_j^-\,,
\end{equation}
which is the helicity operator [for definition of
$\epsilon_{ij}$ see (\ref{meigenv})].

To clarify the algebra structure from the point of view of
space-time transformations, let we rewrite our algebra in
terms of $P_i$ and $J_{vi}$ [see (\ref{realdec})]:

\begin{equation}\label{rdifreal1}
P_i=\cos u\partial_i-\sin u\zeta_i\partial_v-\sin u M_{Vi}\,,
\end{equation}
\begin{equation}\label{rdifreal2}
J_{vi}=\sin u\partial_i+\cos u\zeta_i\partial_v+\cos u M_{Vi}\,,
\end{equation}
which satisfy the commutation relations

$$
{} [P_i,\, J_{jk}]=-\delta_{ij}P_k +\delta_{ik}P_j\,,
$$

$$
{}[J_{vi},\, J_{jk}]=-\delta_{ij}J_{vk} +\delta_{ik}J_{vj}\,,
$$

$$
{}[P_u,\,J_{vi}]=P_i\,,
\qquad
{}[P_i,\,J_{vj}]=\delta_{ij}P_v\,,
$$

$$
{}[P_u,\,P_i]=-J_{vi}\,.
$$
From the expressions
(\ref{difreal1}),(\ref{rdifreal1}, and (\ref{rdifreal2}), it is
clear that the generators

$$
P_u,\quad P_v,\quad P_i\,,
$$
do not leave point $x^\mu=0$ invariant, while rotation generators
(i.e., their orbital parts)

$$
J_{ij}\,,\quad J_{vi}\,,
$$
leave the point $x^\mu=0$ invariant. In other words, the
generators $P_u$, $P_v$, and $P_i$ can be interpreted as curved
counterparts of translation elements of Poincare algebra
while $J_{ij}$ and $J_{vi}$ are counterparts of the Lorentz
subalgebra.  In terms of these generators the first Casimir
operator [see \ref{casimir2}] reads as

\begin{equation}\label{casimir}
Q=2P_uP_v-P_iP_i-J_{vi}J_{vi}\,.
\end{equation}

Now we are ready to discuss two possible definitions of
masslessness  based on conformal algebra. The first definition
is formulated as the following requirements on the wavefunction:

\begin{equation}\label{1def}
[G,\,Q]|\Phi\rangle=0\,,
\end{equation}
where $G$ are all of the elements of conformal algebra.
The second definition actually is based on so-called $sim(3,1)$
algebra, which is the isometry algebra combined with dilatation
(see Ref.\cite{bhans}). Since elements of isometry algebra
commutate with $Q$, the second definition amounts to the
condition

\begin{equation}\label{def2}
[D,\,Q]|\Phi\rangle=0\,,
\end{equation}
which should, strictly speaking, be accompanied by certain things
(see below).  Note that for the case of Minkowski space-time
the second definition describes all massless states provided by
considerations of Poincare algebra representations.  A
surprising fact discovered in Ref.\cite{SIEG} is that the first
definition describes all massless states only for the case of
four dimensional ($d=4$) Minkowski space-time.  Namely, in
Ref.\cite{SIEG} it was demonstrated that for the massless
representations of Poincare algebra in $d>4$ the first
definition leads to so-called degenerate representations
which constitute only a subset of all massless states. This
result for the case of massless representations of an anti-de
Sitter algebra has been generalized in Ref.\cite{met}, i.e., it
seems that this phenomena is an inherent feature of all massless
fields irrespective of the manifolds were they propagate.  Since
we are considering four-dimensional space-time we expect that
the first definition describes all massless representation,
i.e., there are no restrictions on allowed value of spin.
Below, among other things, we will demonstrate that this is
indeed the case.

Let us start with the second definition. Because of the relation
[see (\ref{kk1}),(\ref{vk1}), and (\ref{realdec})]

$$
{} [D,\,Q]=-2Q\,,
$$
we conclude that wave equation for massless fields must read

\begin{equation}\label{feq1}
Q|\Phi\rangle=0\,.
\end{equation}
As is known in four-dimensional space-time  all massless fields
can be described by means of totally symmetric tensor fields,
i.e., in terms of potentials. In this case to provide a complete
description of massless fields the equation of motion (\ref{feq1})
must be supplemented by corresponding gauge transformation and
subsidiary conditions.  Now we are going to establish such
conditions and gauge transformation.

First of all, due to (\ref{difreal1}) and (\ref{difreal2})
one has the following representation for the operator $Q$:

\begin{equation}\label{casrep}
Q=2e_{_U}e_{_V}-e_ie_i
-2\zeta_iM_{Vi}e_{_V}-M_{Vi}M_{Vi}\,,
\end{equation}
where the dual forms $e_{_A}$ are given by (\ref{dforms}).
Then we introduce the gauge transformation

\begin{equation}\label{gtrans1}
\delta|\Phi\rangle=R|\alpha\rangle
\end{equation}
and postulate the following subsidiary conditions:

\begin{eqnarray}
\label{fcon1}
&& \bar{R}|\Phi\rangle=0
\qquad
(\hbox{divergencelessness})\,,
\\ [5pt]
\label{fcon2}
&&
\bar{a}^{{\scriptscriptstyle A}}\bar{a}_{_A}|\Phi\rangle=0
\qquad
(\hbox{tracelessness})\,,
\\ [5pt]
\label{fcon3}
&&
\bar{a}_{_V}^2|\Phi\rangle=0\,,
\end{eqnarray}
where the operators $R$ and $\bar{R}$ are defined by

$$
R\equiv a^A D_A\,,\quad \bar{R}\equiv \bar{a}^A D_A\,,
$$
$$
D_A=e_A^\mu D_\mu\,,
$$
and $D_\mu$ is a Lorentz covariant derivative

$$ D_\mu\equiv
\partial_\mu+\frac{1}{2}\omega_\mu^{AB}M_{AB}\,.
$$
From (\ref{connec}) it is readily seen that
$D_\mu=\partial_\mu-\delta_\mu^u \zeta_iM_{Vi}$ where a relevant
representation for $M_{AB}$ is given by (A.1). The
$|\alpha\rangle$ in (\ref{gtrans1}) is a parameter of gauge
transformation. If the $|\Phi\rangle$ is a degree-$s$
monomial in the oscillator $a_{_A}$ then the $|\alpha\rangle$
is a degree-$(s-1)$ monomial in $a_{_A}$.  We impose on
$|\alpha\rangle$ the following equation of motion:

\begin{equation}\label{gpfeq}
Q|\alpha\rangle=0
\end{equation}
and the subsidiary conditions

\begin{eqnarray}
\label{gpcon1}
&& \bar{R}|\alpha\rangle=0
\qquad
(\hbox{divergencelessness})\,,
\\ [5pt]
\label{gpcon2}
&&
\bar{a}^{{\scriptscriptstyle A}}\bar{a}_{_A}|\alpha\rangle=0
\qquad
(\hbox{tracelessness})\,,
\\ [5pt]
\label{gpcon3}
&& \bar{a}_{_V}|\alpha\rangle=0\,.
\end{eqnarray}
Let us make a few comments on the gauge transformation and
subsidiary conditions above. As to
(\ref{fcon1},\ref{gpcon1}), (\ref{fcon2}), and (\ref{gpcon2})
these are nothing but familiar conditions of divergencelessness
and tracelessness respectively formulated on tangent space
tensor fields.  The conditions  (\ref{fcon3}) and
(\ref{gpcon3}) are less familiar. These can be rewritten in the
following covariant fashion:

\begin{equation}\label{covcon}
R_{AB}\bar{a}^A\bar{a}^B|\Phi\rangle=0\,, \qquad
R_{AB}\bar{a}^AD^B|\alpha\rangle=0\,,
\end{equation}
where $R_{AB}=2\delta_A^U\delta_B^U$ is the Ricci tensor
in tangent space. Note that the second subsidiary condition
from (\ref{covcon}) is equivalent to (\ref{gpcon3}) provided
$\partial_v|\alpha\rangle={\kern-2.2ex \hbox{/}}\,\,0$.

Making use of (\ref{gpfeq})-(\ref{gpcon3})
and commutation relations

$$
{} [R,\,\bar{R}]=Q+2M_{Vi}M_{Vi}\,,
$$

$$
{}
[R,\,Q]=0\,,
\qquad
[\bar{R},\, Q]=0\,,
$$
$$
{} [\bar{a}^{{\scriptscriptstyle A}}\bar{a}_{_A},\,R]
=-2\bar{R}\,,
\quad
{} [a^{\scriptscriptstyle A}a_{_A},\,\bar{R}]=2R\,,
$$
one can make sure that gauge transformation (\ref{gtrans1})
respects the equation (\ref{feq1}) and subsidiary conditions
(\ref{fcon1})-(\ref{fcon3}).

Now we are going to prove the following main
result of this subsection.

{\it Proposition}:  Let the gauge parameter $|\alpha\rangle$
satisfy (\ref{gpfeq})-(\ref{gpcon3}). Then the equation
(\ref{feq1}) and subsidiary conditions
(\ref{fcon1})-(\ref{fcon3}) supplemented by gauge transformation
(\ref{gtrans1}) describe the $|\Phi^{(0)}_{tr}\rangle$ which (1)
depends only on transversal oscillator $a_i$; and (2) satisfies
the equation of motion

\begin{equation}\label{feq5}
\Box_0|\Phi^{(0)}_{tr}\rangle=0
\end{equation}
and tracelessness condition in transversal directions

\begin{equation}\label{trtrac}
\bar{a}_i\bar{a}_i|\Phi_{tr}^{(0)}\rangle=0\,,
\end{equation}
where in (\ref{feq5}) $\Box_0=2e_{_U}e_{_V}-e_ie_i$ is the
second-order operator for massless scalar field

\begin{equation}\label{scaloper}
\Box_0=2\partial_u\partial_v-\zeta^2\partial_v^2-\partial^2\,.
\end{equation}
Note that (\ref{trtrac}) tells us that actually there are
only two polarization degrees of freedom, while (\ref{feq5})
expresses the fact that these physical degrees of freedom satisfy
the same equation of motion as the spinless (scalar) field. In
these respects there is analogy with massless fields in
Minkowski space-time.

{\it Proof}: From (\ref{fcon3}) and (\ref{gpcon3}) it follows

\begin{equation}\label{decom1}
|\Phi\rangle=|\Phi^{(0)}\rangle+a_{_U}|\Phi^{(1)}\rangle\,,
\end{equation}

\begin{equation}\label{decom2}
|\alpha\rangle=|\alpha^{(0)}\rangle\,,
\end{equation}
where generating functions
$|\Phi^{(0)}\rangle$, $|\Phi^{(1)}\rangle$,
and $|\alpha^{(0)}\rangle$ are independent of $a_{_U}$.
Now substituting (\ref{decom1}) into (\ref{feq1}) we get
for $|\Phi^{(0)}\rangle$ and $|\Phi^{(1)}\rangle$
the following equations of motion:

\begin{equation}\label{feq2}
Q|\Phi^{(0)}\rangle+2(\zeta_ia_ie_{_V}
-a_{_V}+a_iM_{Vi})|\Phi^{(1)}\rangle=0\,,
\end{equation}

\begin{equation}\label{feq3}
Q|\Phi^{(1)}\rangle=0\,.
\end{equation}
By substituting (\ref{decom1}) into (\ref{gtrans1})
we get the gauge transformations

\begin{equation}\label{gtrans2}
\delta|\Phi^{(0)}\rangle=(a_{_V}(e_{_U}-\zeta_iM_{Vi})
-a_ie_i)|\alpha^{(0)}\rangle\,,
\end{equation}

\begin{equation}\label{gtrans3}
\delta|\Phi^{(1)}\rangle=e_{_V}|\alpha^{(0)}\rangle\,,
\end{equation}
and by substituting (\ref{decom1}) into
(\ref{fcon1}) and (\ref{fcon2}) we obtain
the constraints

\begin{equation}\label{newcon1}
(\bar{a}_{_U}e_{_V}-\bar{a}_ie_i)|\Phi^{(0)}\rangle
-(e_{_U}-\zeta_iM_{Vi})|\Phi^{(1)}\rangle=0\,,
\end{equation}

\begin{equation}\label{newcon2}
(\bar{a}_{_U}e_{_V}-\bar{a}_ie_i)|\Phi^{(1)}\rangle=0\,,
\end{equation}

\begin{equation}\label{newcon3}
\bar{a}_i\bar{a}_i|\Phi^{(0)}\rangle
+2\bar{a}_{_U}|\Phi^{(1)}\rangle=0\,,
\end{equation}

\begin{equation}\label{newcon4}
\bar{a}_i\bar{a}_i|\Phi^{(1)}\rangle=0\,.
\end{equation}
By using (\ref{gpfeq})-(\ref{gpcon2}) and
(\ref{decom2}) we repeat the procedure above for
$|\alpha^{(0)}\rangle$ and get the following equation of motion
and constraints:

\begin{equation}\label{feq4}
Q|\alpha^{(0)}\rangle=0\,,
\end{equation}

\begin{equation}\label{newgpcon1}
(\bar{a}_{_U}e_{_V}-\bar{a}_ie_i)|\alpha^{(0)}\rangle=0\,,
\end{equation}

\begin{equation}\label{newgpcon2}
\bar{a}_i\bar{a}_i|\alpha^{(0)}\rangle=0\,.
\end{equation}
Now taking into account that $|\Phi^{(1)}\rangle$ and
$|\alpha^{(0)}\rangle$ satisfy the same equations of motion
[see (\ref{feq3}] and (\ref{feq4})) as well as the same
constraints [see (\ref{newcon2}),(\ref{newcon4})
(\ref{newgpcon1}), and (\ref{newgpcon2})] we can, due to
(\ref{gtrans3}), impose the following light cone like gauge:

\begin{equation}\label{gauge}
|\Phi^{(1)}\rangle=0\,.
\end{equation}
By substituting the gauge (\ref{gauge}) into (\ref{newcon3}) and
(\ref{newcon1}) we get

\begin{equation}\label{trac3}
\bar{a}_i\bar{a}_i|\Phi^{(0)}\rangle=0\,,
\end{equation}

\begin{equation}\label{con5}
(\bar{a}_{_U}e_{_V}-\bar{a}_ie_i)|\Phi^{(0)}\rangle=0\,.
\end{equation}
Constraint (\ref{trac3}) tells us that
$|\Phi^{(0)}\rangle$ is a traceless tensor with respect to
transversal indices $i_1,\ldots ,i_s$. The solution to
(\ref{con5}) can be written as

\begin{equation}\label{fincon}
|\Phi^{(0)}\rangle=
\exp(-\frac{e_i}{e_{_V}}\bar{a}_ia_{_V})|\Phi^{(0)}_{tr}\rangle\,,
\end{equation}
where subscript (tr) indicates the fact that
$|\Phi^{(0)}_{tr}\rangle$ depends only on the transversal
oscillator $a_i$. By substituting (\ref{fincon}) into
(\ref{trac3}) we get (\ref{trtrac}), while by substituting
(\ref{fincon}) and (\ref{gauge}) into
(\ref{feq2}) we arrive at (\ref{feq5}). The proposition is
proved.

In conclusion of this subsection let us note that helicity
operator $Q_\lambda$ from (\ref{heloper}) is realized on
$|\Phi^{(0)}_{tr}\rangle$ as follows:

$$
Q_\lambda|\Phi^{(0)}_{tr}\rangle
=\epsilon_{ij}M_{ij}\partial_v|\Phi^{(0)}_{tr}\rangle\,,
$$
and as is easily seen takes two values equal to $\pm \lambda p_v$.

\subsection{Light cone representation
for conformal algebra}

In  this subsection by making use of the first definition we
shall derive the expressions for elements of conformal algebra
acting as differential operators on the solution of (\ref{feq5})
and defined on a surface of initial conditions $u=0$.  In other
words we are going to investigate the complete set of
the equations which follows from (\ref{1def}). As is well known,
the complete set of equations is obtained by performing  double
conformal boost, i.e.,  the complete set of equations is given
by $[G,\,Q]|\Phi\rangle=0$ as well as
$[G,\,[G,\,Q]]|\Phi\rangle=0$.  Note that the set of equations
in question can be rewritten in a form which is manifestly
invariant with respect to conformal algebra:

\begin{equation}\label{confeq}
T_{AB}=0\,,
\end{equation}
where

$$
T_{AB}=\{ J_{AC},\,J_{CB}\}
+\frac{1}{3}\eta_{AB}^{\vphantom{5pt}}J_{CD}J_{CD}\,,
$$
where $\{a,b\}\equiv ab+ba$.
These equations we analyze  on the surface of initial
conditions $u=0$ and make use of light cone like gauge

\begin{equation}\label{lclg}
M_{Vi}=0\,.
\end{equation}
Note that we prefer to pass from wavefunction
$\Phi(u,v,\zeta_i)$ to its Fourier modes with respect to $v$,
i.e.,  in what follows we shall replace operator $\partial_v$ by
$-{\rm i}p_v$.

Again one has the equation of motion (\ref{feq1}),
making use of which as well as (\ref{casrep}) and (\ref{lclg})
we get on-mass shell condition

$$(\partial_u\Phi)\Bigl|_{u=0}=\frac{{\rm
i}}{2p_v}(\partial^2-p_v^2\zeta^2)\Phi(p_v,\zeta)\,.
$$
Now making use of (\ref{link}) and (\ref{oskdec}) we are able to
learn restrictions which impose (\ref{confeq}) on $M_{AB}$,
$\Delta$, and $\kappa_A$.  Because the analysis of
eqs.(\ref{confeq}) is straightforward but very tedious, we
outline the procedure of solution and present the results.

From $T_{+V}=0$ one gets

$$
M_{UV}=\frac{1}{2}(1-\Delta)\,,
$$
while from $T_{+i}=0$ we learn
$$
M_{Ui}=-\frac{{\rm i}}{p_v}M_{ij}\partial_j
+\frac{{\rm i}}{p_v}\partial_i\,.
$$
Making use of these results we derive from equations
$T_{--}=0$, $T_{-i}=0$, and $T_{+-}=0$,

\begin{equation}
\kappa_{_V}=0\,,  \qquad
\kappa_i=0\,,  \qquad
\Delta=1\,,
\end{equation}
respectively, while from $T_{+i}=0$ one gets

$$
\kappa_{_U}=\frac{{\rm i}}{2p_v}M^2\,,
\qquad M^2\equiv M_{ij}M_{ij}\,.
$$
Now collecting all results above we get the following
light cone representation for the generators of conformal group
acting on the wavefunction of plane wave geometry massless fields
$\Phi(p_v,\zeta)$:

$$
(\pm){\bf P}_u=\frac{{\rm i}}{2p_v}\partial^2\,,
\qquad
(\pm){\bf P}_v=-{\rm i}p_v\,,
\qquad
(\pm){\bf P}_i=\partial_i\,,
$$

\begin{eqnarray}
&&{\bf J}_{ij}=
-\zeta_i \partial_j+\zeta_j \partial_i + M_{ij}\,,
\nonumber                               \\[5pt]
&& {\bf J}_{vi}
=-{\rm i}\zeta_ip_v\,,
\nonumber                               \\ [5pt]
&& {\bf J}_{uv}=-{\rm i}vp_v\,,
\nonumber                               \\ [5pt]
&& {\bf J}_{ui}=v\partial_i
+\frac{{\rm i}}{2p_v}\zeta_i\partial^2
-\frac{{\rm i}}{p_v}
M_{ij}\partial_j\,,
\nonumber                                \\
\label{jkgen} &&               \\[-3pt]
&& (\pm){\bf K}_u=v{\bf D}+
\frac{{\rm i}}{4p_v}\zeta^2\partial^2
-\frac{{\rm i}}{p_v}\zeta_iM_{ij}\partial_j+
\frac{{\rm i}}{2p_v}M^2\,,
\nonumber                             \\ [5pt]
&& (\pm){\bf K}_v=-\frac{{\rm i}}{2}\zeta^2p_v\,,
\nonumber                                 \\ [5pt]
&& (\pm) {\bf K}_i=-\zeta_i{\bf D}+\frac{1}{2}\zeta^2\partial_i
+M_{ij}\zeta_j\,,
\nonumber  \\ [5pt]
&& {\bf D}=-{\rm i}vp_v+\zeta_j\partial_j+1\,.
\nonumber
\end{eqnarray}
Note that while writing the final representation (\ref{jkgen})
we have assigned $+$ and $-$ to particles and antiparticles,
respectively, and have put $p_v>0$ for both particles and
antiparticles.  It should be
emphasized that all analysis above is sensitive to space-time
dimension because from the equation $T_{ij}=0$ it follows

\begin{equation}\label{degcon}
\{M_{ik},\,M_{jk}\}=\delta_{ij}M^2\,.
\end{equation}
It is the equation (\ref{degcon}) that leads to degenerate
representation in higher space time dimensions $d>4$
(see Refs.\cite{SIEG} and \cite{met}).
Fortunately, in four dimensions, due to relations

\begin{equation}\label{meigenv}
M_{ij}={\rm i}\epsilon_{ij}\lambda\,,
\qquad
\epsilon_{ij}=-\epsilon_{ji}\,,
\qquad
\epsilon_{12}=1\,,
\end{equation}
the equation (\ref{degcon}) does not impose of any restriction
on helicity value $\lambda$.

\subsection{Energy lowest value}

In this subsection  we are going to demonstrate one interesting
phenomena of massless fields in plane wave geometry.  Namely we
would like to show explicitly that (1) energy operator ${\rm
i}P_u$ (${\rm i}P_u\Phi={\rm i}\partial_u\Phi$) is bounded from
below (above) for particles (antiparticles) and (2) it takes
only discrete values.

As we have shown physical degrees of freedom of
massless arbitrary spin fields satisfy the equation

\begin{equation}\label{eq2}
(2\partial_u\partial_v-\zeta^2\partial_v^2
-\partial_i^2)\Phi=0\,.
\end{equation}

First of all we introduce a radial variable $\eta$ and angle
$\phi$ by

\begin{equation}\label{subst}
\zeta=\frac{\sqrt{\eta}}{2\sqrt{|p_v|}}e^{{\rm i}\phi}\,.
\end{equation}
Note that in (\ref{subst}) we use the complex variable $\zeta$
given by (\ref{complex}).  Then we pass to Fouirer modes with
respect to $u$, $v$, and $\phi$, i.e., we put

$$
\partial_u\Phi=-{\rm i}p_u\Phi\,,\quad
\partial_v\Phi=-{\rm i}p_v\Phi\,,\quad
\partial_\phi\Phi={\rm i}k\Phi\,,
$$
$k=0,\pm1,\pm2\ldots$, and use the substitution

\begin{equation}\label{sub2}
\Phi=e^{-\eta/2}\eta^{|k|/2}\omega(\eta)\,.
\end{equation}
As a result, one has a differential equation for degenerate
hypergeometric function

\begin{equation}\label{hupeq}
\eta\omega^{\prime\prime}
+(\gamma-\eta)\omega^\prime-\alpha\omega=0\,,
\end{equation}
where in (\ref{hupeq}) prime indicates partial derivative with
respect to $\eta$ while $\gamma$ and $\alpha$ are given by

\begin{equation}\label{pu}
\gamma=|k|+1\,,
\qquad
\alpha=
\frac{1}{2}(-p_u{\rm sign} p_v+|k|+1)\,,
\end{equation}
where ${\rm sign}\, p_v =1 (-1)$ for $p_v>0(p_v<0)$.
Solution to (\ref{hupeq}) is $\omega=F(\alpha,\gamma,\eta)$.
It is straightforward now verify (see Ref.\cite{ablsteg}) that
in order for $\Phi$ [see \ref{sub2}] to be square-integrable
one should take
$\alpha=-n$, where $n=0,1,\ldots$.  For such $\alpha$ the
$\omega$ is nothing but the generalized Laggere polynomial
$\omega=L_n^{(|k|)}(\eta)$.  Now taking into account (\ref{pu}),
we arrive at the following spectrum for $p_u$:

\begin{equation}\label{spectr}
p_u={\rm sign} p_v(2n+|k|+1)\,.
\end{equation}
Note that with to respect the dimensionfull parameter $F$ we
should multiply the rhs of (\ref{spectr}) by $\sqrt{F}$.  The
$p_v>0$ corresponds to particles while $p_v<0$ corresponds to
antiparticles.  Note that in contrast to particles in
Minkowski space-time, the $p_u$ does not reach zero value:

$$
p_v>0;\qquad {\rm min}\,p_u=1\,,\quad
$$
$$
p_v<0;\qquad {\rm max}\,p_u=-1\,.
$$
In this respect plane wave geometry massless particles are
similar to anti-de Sitter massless particles (see
Refs.\cite{F1}-\cite{BRFR}).
The analogy to de-Sitter particles can
be drawn further. First of all there exists vacuum
state $|vac\rangle$ defined by $P_i^-|vac\rangle=0$. Second,
level $n$ states can be obtained by operating with $P_i^+$ on
the vacuum: $P_{i_1}^+\ldots P_{i_n}^+|vac\rangle$. Note that
making use of equation $Q|vac\rangle=0$ and (\ref{casimir}) we
immediately obtain the lowest energy value mentioned above.

\setcounter{equation}{0}

\section{Ladder Representation}

In this section we would like to make contact with
Ref.\cite{MacTod}. In Ref.\cite{MacTod} (see also Ref.\cite{F2})
it was demonstrated that a chain of massless representations of
Poincare algebra consisting of all helicity states
$\lambda=0,\pm 1/2,\pm 1, \ldots$ (where every state appears
just once) can be embedded into a ladder representation of
$u(2,2)$ algebra.  We will demonstrate that the same chain of
plane wave geometry massless states can also be embedded into
the same ladder representation. As result we will able to show
explicitly the manner in which Minkowski and plane wave geometry
massless fields are related to each other.

A second long-term motivation of our interest in ladder
representation comes from recent progress in the problem of
massless higher spin fields dynamics (\cite{VAS1}).  Namely,
for the case of anti-de Sitter space-time in Ref.(\cite{VAS1})
it was demonstrated that to construct self-consistent
interaction of massless higher spin fields it is necessary,
among other things, to introduce the chain anti-de Sitter
massless fields which consists of every spin just once. Due to
that, it is strongly believed that ladder representations can
get interesting applications in future studies  of massless
higher spin fields interactions and deserve further
investigation.

Let us briefly describe how the ladder
representation is constructed (for details see
Refs.\cite{MacTod} and \cite{StTod}).  Define the operator-valued
four-component spinor $\varphi=(\varphi^\alpha)$ and
$\tilde{\varphi}=\varphi^*\beta=(\tilde{\varphi}_\alpha)$,
$\alpha=1,2,3,4$, and impose the commutation relations

\begin{equation}\label{funcr}
{}
[\varphi^\alpha,\,\tilde{\varphi}_\beta]=\delta_\beta^\alpha\,,
\end{equation}
where

\begin{equation}\label{betamat}
\beta=\pm \gamma_0
\end{equation}
and $\gamma_\mu$ are Dirac
$\gamma$-matrices subjected as usual to relation

$$
\{\gamma_\mu,\gamma_\nu\}=2\eta_{\mu\nu}\,,
\qquad
\eta_{\mu\nu}={\rm diag}(1,-1,-1,-1)\,.
$$
The signs $+$ and $-$ in (\ref{betamat}) and below correspond to
representations which are associated with particle or
antiparticle, respectively. Now the ladder representation of
$u(2,2)$ is defined by

$$
J_{AB}=\tilde{\varphi}\Sigma_{AB}\varphi\,,
\qquad
C_1=\tilde{\varphi}\varphi\,,
$$
where
$$
\Sigma_{AB}=\{ \Sigma_{\mu\nu},\Sigma_{5\mu},
\Sigma_{6\mu},\Sigma_{56}\}\,,
$$
$$
\Sigma_{\mu\nu}=\frac{1}{4}[\gamma_\mu,\,\gamma_\nu]\,,
\qquad
\Sigma_{56}=\frac{{\rm i}}{2}\gamma_5\,,
$$
$$
\Sigma_{5\mu}=-\frac{{\rm i}}{2}\gamma_\mu\,,
\qquad
\Sigma_{6\mu}=\frac{1}{2}\gamma_\mu\gamma_5\,.
$$
Note that $\Sigma_{AB}$ satisfy the same commutations relations
as $J_{AB}$. We use the following representation for
$\gamma$-matrices:

$$
\gamma_\mu=\left(
\begin{array}{ll}
0 & \bar{\sigma}_\mu\\
\sigma_\mu & 0
\end{array}
\right)\,,
\qquad
\sigma_\mu=(1,\sigma_i)\,,
\qquad
\bar{\sigma}_\mu=(1,-\sigma_i)\,,
$$
where $\sigma_i$, $i=1,2,3$, are usual Pauli matrices. Note
that $\sigma_{u,v}$ are given by
$\sigma_u=(\sigma_0+\sigma_3)/\sqrt{2}$ and
$\sigma_v=(\sigma_0-\sigma_3)/\sqrt{2}$.
The $\gamma_5$ is defined
by $\gamma_5=\gamma_0\gamma_1\gamma_2\gamma_3$.

To carry out the reduction we should rewrite the
elements of the $u(2,2)$ algebra as differential operators acting
on functions of $p_v$ and $\zeta_i$. To do that let us consider
the space of Hillbert space of finite norm  functions where the
norm is defined by

\begin{equation}\label{norm}
(f,g)=\int d^2z_1 d^2 z_2 \overline{f(z_1,z_2)}g(z_1,z_2)\,.
\end{equation}
Now we choose the following representation for
operator-valued four-component spinor $\varphi$

\begin{equation}\label{2spin}
\varphi=\left(
\begin{array}{c}
\lambda\\
\pm \tau
\end{array}\right)\,,
\qquad
\lambda=\left(
\begin{array}{c}
\partial_{z_1}\\
\bar{z}_2
\end{array}\right)\,,
\qquad
\tau=\left(
\begin{array}{c}
\bar{z}_1\\
\partial_{z_2}
\end{array}\right)\,,
\end{equation}
where we introduce two-component spinors $\lambda=(\lambda^a)$
and $\tau=(\tau^a)$, a=1,2, which satisfy the commutation
relations

$$
{}[\lambda^a,\, \tau_b^*]=\delta_b^a\,,\qquad
{}[\tau^a,\, \lambda_b^*]=\delta_b^a\,.
$$
Due to (\ref{norm}) we have the following involution rules
$(\partial/\partial z)^*=-\partial/\partial z$ and
$z^*=\bar{z}$,
i.e.,
$$
\tilde{\varphi}=(\tau^*,\pm \lambda^*)\,,\qquad
\lambda^*=(-\partial_{\bar{z}_1}, z_2)\,,
\qquad
\tau^*=(z_1,-\partial_{\bar{z}_2})\,.
$$
Making use of (\ref{Lorsix}) and (\ref{2spin}) we derive the
following representation for elements of $u(2,2)$ algebra in
Lorentz basis

$$ (\pm){\bf P}_\mu
=-\frac{{\rm i}}{\sqrt{2}}\lambda^*\sigma_\mu\lambda\,,\qquad
(\pm){\bf K}_\mu
=-\frac{{\rm i}}{\sqrt{2}}\tau^*\bar{\sigma}_\mu\tau\,,
$$
$$
{\bf J}_{\mu\nu}=\lambda^*\sigma_{\mu\nu}\tau
+\tau^*\bar{\sigma}_{\mu\nu}\lambda\,,\qquad
{\bf D}=\frac{1}{2}(\tau^*\lambda-\lambda^*\tau)\,,
$$
$$
C_1=\lambda^*\tau+\tau^*\lambda\,,
$$
which is very convenient for practical calculations. Making use
of this representation we get the following expressions for
generators:

$$
\begin{array}{l}
(\pm){\bf P}_u
={\rm i}\partial_{z_1}\partial_{\bar{z}_1}\,,
\\ [3pt]
(\pm){\bf P}_v=-{\rm i}z_2\bar{z}_2 \,,
\\ [3pt]
(\pm){\bf P}_\zeta=-{\rm i}z_2\partial_{z_1}\,,
\\ [3pt]
(\pm){\bf P}_{\bar{\zeta}}
={\rm i}\bar{z}_2\partial_{\bar{z}_1}\,,
\end{array}
\quad
\begin{array}{l}
(\pm){\bf K}_u
={\rm i}\partial_{z_2}\partial_{\bar{z}_2}\,,
\\ [3pt]
(\pm){\bf K}_v=-{\rm i}z_1\bar{z}_1\,,
\\ [3pt]
(\pm){\bf K}_\zeta=-{\rm i}\bar{z}_1\partial_{\bar{z}_2}\,,
\\ [3pt]
(\pm){\bf K}_{\bar{\zeta}} ={\rm i}z_1\partial_{z_2}\,,
\end{array}
\quad
\begin{array}{l}
{\bf J}_{u\zeta}=-\partial_{z_1}\partial_{\bar{z}_2}\,,
\\ [3pt]
{\bf J}_{u\bar{\zeta}}=\partial_{\bar{z}_1}\partial_{z_2}\,,
\\ [3pt]
{\bf J}_{v\zeta}=-\bar{z}_1z_2\,,
\\ [3pt]
{\bf J}_{v\bar{\zeta}}=z_1 \bar{z}_2\,,
\end{array}
$$
$$
{\bf J}_{uv}=-\frac{1}{2}(
z_1\partial_{z_1}+
z_2\partial_{z_2}+
\partial_{\bar{z}_1}\bar{z}_1+
\partial_{\bar{z}_2}\bar{z}_2)\,,
$$
$$
{\bf D}=\frac{1}{2}(
z_1\partial_{z_1}-
z_2\partial_{z_2}+
\partial_{\bar{z}_1}\bar{z}_1-
\partial_{\bar{z}_2}\bar{z}_2)\,,
$$
$$
{\bf J}_{\zeta\bar{\zeta}}
=\frac{1}{2}(z_1\partial_{z_1}-z_2\partial_{z_2}
-\partial_{\bar{z}_1}\bar{z_1}
+\partial_{\bar{z}_2}\bar{z}_2)\,,
$$
$$
C_1=z_1\partial_{z_1}
+z_2\partial_{z_2}
-\partial_{\bar{z}_1}\bar{z_1}
-\partial_{\bar{z}_2}\bar{z_2}\,.
$$
Now we link the variables $p_v$ and $\zeta_i$ with $z_1$ and
$z_2$ as follows:

$$
\zeta={\rm i}\frac{z_1}{z_2}\,,\qquad
p_v=z_2\bar{z}_2\,,\qquad
\phi=\frac{{\rm i}}{2}\ln \frac{\bar{z}_2}{z_2}\,,
$$
$$
p_v>0\,,\qquad \phi \in [0,2\pi]\,,
$$
where we again use the complex variable $\zeta$ given by
(\ref{complex}) and introduce an additional variable $\phi$.
Note that in contrast to Minkowski space-time (see
Ref.\cite{MacTod}), where $z_i$ are linked with lightlike
momentum $p_\mu$ in our case $z_i$ are linked with momentum
$p_v$ as well as space coordinates $\zeta_i$.  It is clear that
the function $f(z_1,z_2)$ can be rewritten as a certain function
of new variables $\Phi(p_v,\zeta_i,\phi)$:

$$
f(z_1,z_2)=\Phi(\zeta_i, p_v, \phi)\,.
$$
The corresponding measure from (\ref{norm}) in terms of new
variables is given by

$$
d^2z_1d^2z_2=p_vd^2\zeta dp_v d\phi\,.
$$
Making use of chain rules

$$
\partial_{z_1}
=\frac{{\rm i}}{z_2}\partial_\zeta\,,
\qquad
\partial_{z_2}
=\frac{1}{z_2}(-\zeta\partial_\zeta+{\rm i}p_v v -
\frac{{\rm i}}{2}\partial_\phi)
$$
and their analog for $\partial/\partial \bar{z}$ one gets the
following representations for generators acting on the functions
$\Phi(p_v,\zeta_i)$:

\begin{eqnarray}
&& {\bf J}_{ij}=-\zeta_i \partial_j+\zeta_j \partial_i
+\frac{1}{2}\epsilon_{ij}\partial_\phi\,,
\nonumber                                 \\[5pt]
&& {\bf J}_{vi}=-{\rm i}\zeta_ip_v \,,
\nonumber                                \\ [5pt]
&& {\bf J}_{uv}=-{\rm i}vp_v \,,
\nonumber                             \\ [5pt]
&& {\bf J}_{ui}=v\partial_i
+\frac{{\rm i}}{2p_v}\zeta_i\partial^2
-\frac{{\rm i}}{2p_v}\epsilon_{ij}\partial_j\partial_\phi\,,
\nonumber \\
\label{conphi}     &&                              \\ [-3pt]
&& (\pm){\bf K}_u=v{\bf D}+
\frac{{\rm i}}{4p_v}\zeta^2\partial^2
-\frac{{\rm i}}{2p_v}\zeta_i\epsilon_{ij}\partial_j\partial_\phi+
\frac{{\rm i}}{4p_v}\partial_\phi^2\,,
\nonumber                           \\ [5pt]
&& (\pm){\bf K}_v=-\frac{{\rm i}}{2}\zeta^2p_v\,,
\nonumber                              \\ [5pt]
&& (\pm){\bf K}_i=-\zeta_i{\bf D}+\frac{1}{2}\zeta^2\partial_i
+\frac{1}{2}\epsilon_{ij}\zeta_j\partial_\phi\,,
\nonumber                                     \\ [5pt]
&& {\bf D}=-{\rm i}vp_v+\zeta_j\partial_j+1\,,
\nonumber
\end{eqnarray}

$$
C_1=-{\rm i}\partial_\phi-2\,,
$$
where in the last expressions we go back to $\zeta_i$
[see (\ref{complex})]. Note that for ${\bf P}_\mu$ the results
are the same as in (\ref{jkgen}).
Now expand $\Phi(p_v,\zeta_i,\phi)$ in
Fourier series with respect to $\phi$ over the interval
$[0,2\pi]$:

\begin{equation}\label{irrcom}
\Phi(p_v, \zeta_i,\phi) =\frac{1}{\sqrt{2\pi}}
\sum_{\lambda=0,\pm\frac{1}{2},\pm1,\ldots}
\Phi_\lambda(p_v, \zeta_i)e^{2{\rm i}\lambda\phi}\,.
\end{equation}
Note that each term under sum in (\ref{irrcom})
transforms into an irreducible representation of $u(2,2)$ for
which $C_1$ is diagonal and takes the values

\begin{equation}\label{helsp}
C_1=2(\lambda-1)\,,
\qquad
\lambda=0,\pm\frac{1}{2},\pm1,\pm\frac{3}{2},\ldots \,,
\end{equation}
where

\begin{equation}\label{phieig}
\partial_\phi=2{\rm i}\lambda\,.
\end{equation}
By substituting (\ref{phieig}) into (\ref{conphi})
and (\ref{meigenv}) into (\ref{jkgen}) we take one and
the same expression. Thus we have proved completely that by
decomposing the ladder representation into irreducible
representations of the $u(2,2)$ algebra, we get a chain of
plane wave geometry massless states with helicities
displayed in (\ref{helsp}) where each state
is included just once.
The corresponding norm is also decomposed into direct sum as

$$
(\Phi,G)=\sum_\lambda \int p_vdp_v d^2\zeta
\overline{\Phi_\lambda(p_v,\zeta)}G_\lambda(p_v,\zeta)\,.
$$

At the end of this section we would like to demonstrate how
Minkowski and plane wave geometry massless fields
are related to each other.
A reason for such an interrelation comes from the fact
that chains of Minkowski as well as plane wave geometry massless
fields realize the same commutation relations (\ref{funcr}).
To demonstrate such interrelation explicitly
let us recall one of the results of Ref.(\cite{MacTod}).
In Ref.(\cite{MacTod}) it was shown that for the case
of Minkowski massless fields it is necessary to choose the
following representation for operator-valued four-component
spinor $\varphi_{Min}^{\vphantom{5pt}}$

\begin{equation}\label{2spinMin}
\varphi_{Min}^{\vphantom{5pt}}=\left(
\begin{array}{c}
\displaystyle{
\lambda_{Min}^{\vphantom{5pt}}           } \\ [5pt]
\displaystyle{
\pm \tau_{Min}^{\vphantom{5pt}}       }
\end{array}\right)\,,
\qquad
\lambda_{Min}^{\vphantom{5pt}}=\left(
\begin{array}{l}
\displaystyle{
\bar{z}_1         }\\ [5pt]
\displaystyle{
\bar{z}_2             }
\end{array}
\right)\,,\qquad
\tau_{Min}^{\vphantom{5pt}}=\left(
\begin{array}{l}
\displaystyle{
\partial_{z_1}               }\\ [5pt]
\displaystyle{
\partial_{z_2}             }
\end{array}
\right)\,.
\end{equation}
Please compare with our choice for the case of plane wave
massless fields (\ref{2spin}). Now what needs to be demonstrated
is that  $\varphi$ and $\varphi_{Min}$ are related by
unitary transformation

$$
\varphi=U\varphi_{Min}^{\vphantom{5pt}}\,,
$$
where the matrix $U\in U(2,2)$--group, i.e., it should satisfy

$$
U^*\beta U=\beta\,.
$$
Our statement is that such a intertwine matrix exists and
is given by

$$
U=\left(
\begin{array}{clcl}
0 & 0 & \pm1 & 0\\
0 & 1 & 0 & 0\\
\pm1 & 0 & 0 & 0\\
0 & 0 & 0 & 1\\
\end{array}
\right)\,.
$$

\section{Conclusion}

We have constructed free equations of motion for
plane wave geometry massless fields of arbitrary spin.  We
believe that this result can get interesting
development and applications.  Let us briefly outline the
following obvious applications.  First of all, the equations in
question could be easily generalized to five-dimensional plane
wave geometry.  Therefore, making use of the procedure of
dimensional reduction, one could get a description of
four-dimensional massive arbitrary spin fields. From this point
of view this result can get applications in systematical
analysis of string massive states spectrum.  Another interesting
potential application our result can get is the problem of
introducing an interaction into the theory of massless higher
spin fields. Note that for the case of anti-de Sitter geometry
this problem, at least at classical level, has been solved
recently in Ref.(\cite{VAS1}). One can speculate that for the
case of plane wave geometry higher massless spin fields this
problem also has a positive solution.  Note that in
Ref.(\cite{VAS1}) it was demonstrated that it is the spectrum of
anti-de Sitter massless states related to the $u(2,2)$ algebra
ladder representation that leads to self-consistent interaction
of anti-de Sitter massless higher spin fields. Therefore it is
strongly believed that our results related to ladder
representations can also get interesting applications in the
problem of interaction of plane wave geometry massless higher
spin fields.  Because detailed investigation of these
questions is too involved, we hope to study them in future
publications.

\section*{Acknowledgements}

The author wishes to thank M.A. Vasiliev for useful suggestions.
This work was supported in part by INTAS, Grant No.94-2317,
by the Russian Foundation for Basic Research,
Grant No.96-01-01144, and by the NATO Linkage, Grant No.931717.

\section*{Appendix A. Spin part of conformal algebra}

For the case of the vector field ($s=1$) the Lie derivative

$$ {\cal L}_\xi \Phi^{\mu_1,\ldots\mu_s}
=\xi^\nu\partial_\nu \Phi^{\mu_1\ldots\mu_s}
-\sum_{k=1}^s
\Phi^{\mu_1\ldots\nu\ldots\mu_s}\partial_\nu
\xi^{\mu_k}
$$
can be rewritten in terms of $\Phi^A$ [see (\ref{trans})] as
follows:

$$
{\cal L}_\xi\Phi^A=\xi^\mu\partial_\mu\Phi^A
-e_{_B}^\mu\Phi^B\partial_\mu\xi^A
+\delta_V^AF(u)(\xi^U\zeta^i \Phi^i
-\Phi^U\zeta^i\xi^i)\,,
$$
where we introduce $\xi^A={\bf e}_\mu^A\xi^\mu$.
Now it is clear how the last expression can be
generalized to the case of arbitrary rank tensor field.
Making use of generating function [see \ref{genfun}]
the Lie derivatives can be written as

$$
{\cal L}_\xi|\Phi\rangle=G|\Phi\rangle\,,
$$
where
$$
G=\xi^\mu\partial_\mu
+e_A^\mu \partial_\mu\xi^B a_B\bar{a}^A
+a_{_V}\zeta^iF(u)(-\xi^U\bar{a}_i +\xi^i\bar{a}_{_V})\,.
$$
The first term on the rhs corresponds to the orbital part
and shall be denoted by $l(G)$, while the rest is called
spin part and shall be denoted by $s(G)$.
Now we are going to rewrite $s(G)$ in a more invariant manner,
namely in terms of $M_{AB}$ [see (\ref{Lorenal})] which, in a
space of totally symmetric tensor fields (\ref{genfun}),
has the following representation:

$$
M_{_{AB}}=-a_{_A}\bar{a}_{_B}+a_{_B}\bar{a}_{_A}\,.
\eqno({A.1})
$$
Making use of
(\ref{conkileq}) one can make sure that the following relations
hold true:
$$
\partial_v\xi^i=\partial_i\xi^U\,,
\qquad
e_U^\mu\partial_\mu\xi^V=-F(u)\zeta^i\xi^i\,,
\qquad \partial_v\xi^U=0\,,
$$
and
$$
-e_U^\mu\partial_\mu\xi^i+\partial_i\xi^V=\zeta^i\xi^U\,,
\qquad
e_U^\mu\partial_\mu\xi^U+\partial_v\xi^V=\frac{1}{2}(\nabla
\xi)\,.
$$
With the help of these relations and (A.1), the expression
for $s(G)$ can be rewritten as

\begin{eqnarray*}
s(G)&=&-\partial_v\xi^V M_{VU}
+(\partial_i\xi^V-F(u)\zeta^i\xi^U)M_{Vi}
+\partial_v\xi^i M_{Ui}                   \\
&+&\frac{1}{4}(\nabla\xi)(\Delta
+2M_{VU})-\frac{1}{2}\partial_i\xi_jM_{ij}\,,
\end{eqnarray*}
where
$$
\Delta=2a_{_V}\bar{a}_{_U}-a_i\bar{a}_i\,.
\eqno{(A.2)}
$$
Note that while deriving $s(G)$ we taken into account only
(\ref{conkileq}). Taking into account equations for isometry
Killing vector $\nabla_\mu\xi_\nu+\nabla_\nu \xi_\mu=0$ we arrive
at (\ref{spinp1}). It is clear that the last
representation for $s(G)$ is valid not only for totally
symmetric tensor fields but for those of arbitrary
symmetry as well.

\section*{Appendix B. Superalgebra osp(2,1)}

In this Appendix we prove the following interesting statement:
if $y_\alpha$ and $Y_a$ are independent solutions of the
equations (\ref{soesc}) and (\ref{toe}) respectively then the
algebra (\ref{supa1}), where the brackets are
defined by (\ref{com1}), is isomorphic to
superalgebra $osp(2,1)$. To prove the statement we choose
the following solutions of Eq.(\ref{toe}):
$$
Y_1=y_1^2\,,\quad Y_2=y_2^2\,,\quad Y_3=y_1y_2\,.
\eqno{(B.1)}
$$
For such a choice of $Y_a$ the determinant of Eq.(\ref{toe})
is equal to $-2W^3$, where $W$ is a
determinant of (\ref{soesc}): $W=y_1y_2^\prime-y_2y_1^\prime$.
Note that $W=const$.
Because $y_\alpha$ form independent solutions of (\ref{soesc}),
i.e.,  $W={\kern-2.2ex/}\,\,0$,  the $Y_a$ in (A.1) also form
independent solutions of Eq.(\ref{toe}).  Now the
statement can be proved by direct calculation. Actually, making
use of (B.1) and (\ref{com1}), it is easy to derive the following
commutation relations:

$$
{}  \{J_0,\,J_\pm\}=\pm J_\pm\,,\quad \{J_+,\,J_-\}=-2J_0\,,
\eqno{(B.2)}
$$
$$
\{J_\pm,\,R_\mp\}=\mp R_\pm\,,
\quad
\{J_0,\,R_\pm\}=\pm\frac{1}{2}R_\pm\,,
\eqno{(B.3)}
$$
$$
\{R_\pm,\,R_\pm\}=J_\pm\,,
\quad
\{R_+,\,R_-\}=J_0\,,
\eqno{(B.4)}
$$
where the notation $J_0=Y_3/W$, $J_+=Y_2/W$,
$J_-=Y_1/W$, $R_+=y_2$ and $R_-=y_1$ has been used. The
relations (B.2)-(B.4) are nothing but
the commutation relations of $osp(2,1)$ superalgebra. The
statement is proved.

\end{document}